\documentclass[preprint,times]{elsarticle}
\title{Multilinear Hirota operators and integrability}
\author[a,b]{Il'in I.A.}
\ead{}
\author[a,b]{Noshchenko D.S.}
\ead{}
\author[a,b]{Perezhogin A.S.}	
\ead{}
\address[a]{Institute of Cosmophysical Research and Radio Wave Propagation of the Far Eastern Branch of Russian Academy of Science, 684034 Kamchatka region, Elizovskiy district, Paratunka, Mirnaya str., 7.}
\address[b]{Vitus Bering Kamchatka State University, 683032, Kamchatka region, Elizovskiy district, Petropavlovsk-Kamchatsky, Pogranichnaya str., 4.}

\begin{document}
\maketitle

\begin{abstract}
We consider multilinear Hirota operators and establish some basic properties.
Next we investigate two special integrable cases, introduced in \cite{Hiet1}, which can be derived
from the bilinear KP-type hierarchies. Finally we discuss a higher order generalizations.
\end{abstract}

\section{Introduction}

Hirota's bilinear derivative \cite{Ath1,Hiet1,Hiet2,JM1} is a powerful tool for construction of integrable nonlinear
equations. For instance, the KdV equation
\begin{equation}
\label{eq:KdV}
u_t + 6 u u_x + u_{xxx} = 0, u(x,t)=\frac{\partial^2}{\partial x^2}\log f(x,t) 
\end{equation}
transforms to bilinear form
\begin{equation}
D_x(D_x^3+D_t) \ f \cdot f = 0,
\end{equation}
where 
\begin{equation}\label{D}
D_{x}^mD_{t}^n=\left(\partial_x - \partial_{x'}\right)^m\left(\partial_t - \partial_{t'}\right)^n \ f(x,t) \cdot f(x',t')|_{x' = x,t'=t}
\end{equation}
or, equivalently,
\begin{equation}
D_{x}^m \ f \cdot f=\frac{d^m}{dy^m}f(x-y)f(x+y)\vline_{y=0}
\end{equation}
Any bilinear equation $G(D_x,D_t)f\cdot{}f=0$, where $G$ is a polynomial, has at least 2SS. 
But 3SS condition is much more stronger. For instance, among the equations
\begin{equation}
(D_xD_t+D_x^n) \ f\cdot{}f=0
\end{equation}
only $n=4,n=6$ has 3SS (KdV and SK equations). We name $n=7$ as ``new'' and $n=9$ as ``tail'' equations.

Here are some integrable bilinear equations, listed in \cite{Hiet1}:
\begin{eqnarray}
(D_x^4-4D_xD_t+3D_y^2) \ f \cdot f=0, \\
(D_x^3D_t+aD_x^2+D_tD_y) \ f \cdot f=0, \\
(D_x^4-D_xD_t^3+aD_x^2+bD_xD_t+cD_t^2) \ f \cdot f=0, \\
(D_x^6+5D_x^3D_t-5D_t^2+D_xD_y) \ f \cdot f=0.
\end{eqnarray}

\section{Multilinear Hirota operators}

Consider k-linear gauge invariant Hirota operator, introduced in \cite{Hiet1}, \cite{Ath1}
\begin{equation}
\left(H_{k,l}^{m}\right)_x=\left(\partial_{x_1}+\gamma_1\partial_{x_2}+\dots+\gamma_{k-1}\partial_{x_k}\right)^m \  f(x_1) \cdot \dots \cdot f(x_k)|_{x_i = x,i=1\dots k}
\end{equation}
$\gamma$ - primitive kth root of unity. For $k=3$ we have
\begin{eqnarray}
T^3=H_{3,1}=\partial_1+e^{\frac{2}{3}i\pi}\partial_2+e^{\frac{4}{3}i\pi}\partial_3 \\
T^{*3}=H_{3,2}=\partial_1+e^{\frac{4}{3}i\pi}\partial_2+e^{\frac{2}{3}i\pi}\partial_3
\end{eqnarray}
For example, Chazy 3rd order integrable equation can be written as a single trilinear determinant
or as a pair of bilinear equations \ref{Hiet1}.
\begin{equation}
 \left| \begin {array}{ccc} f''&f'&f
\\\noalign{\medskip}f'''&f''&f'\\\noalign{\medskip}f''''
&f'''&f''\end {array} \right|=(T^6-T^3 T^{*3}) \ f\cdot{}f\cdot{}f=
\end{equation}
\begin{equation}
=f \left(D_x^3(D_{\tau'}+D_x^3) \ f \cdot f\right),
\end{equation}
with $D_x(D_{\tau'}+D_x^3) \ f \cdot f =0$.

Denote $Q_1^lQ_2^mQ_3^n=Q^{l,m,n}$ -- quadrilinear Hirota operator. In a similar way, the 4th order determinant
\begin{equation}
 \left| \begin {array}{cccc} {\it f'''}&{\it f''}&{\it f'}&f
  \\\noalign{\medskip}{\it f''''}&{\it f'''}&{\it f''}&{\it f'}
\\\noalign{\medskip}{\it f'''''}&{\it f''''}&{\it f'''}&{\it f''}
\\\noalign{\medskip}{\it f''''''}&{\it f'''''}&{\it f''''}&{f'''} \end {array} \right|=
\end{equation}
\begin{equation}
\frac{1}{5\cdot 7 \cdot 2^{14}}\left(45\,Q^{{9,2,1}}+38\,Q^{{7,2,3}}-7\,Q^{{5,2,5}}-8\,Q^{{8,4,0}}+72\,Q^{{
2,8,2}}-140\,Q^{{3,6,3}}\right),
\end{equation}
which is a weighted sum of the following operator determinants
\begin{equation}
\left| \begin {array}{ccc} {Q_{{1}}}^{9}&0&{Q_{{1}}}^{5}
\\\noalign{\medskip}0&{Q_{{2}}}^{2}&0\\\noalign{\medskip}{Q_{{3}}}^{5}
&0&Q_{{3}}\end {array} \right| , \left| \begin {array}{ccc} 0&{Q_{{1}}
}^{8}&{Q_{{1}}}^{2}\\\noalign{\medskip}{Q_{{2}}}^{8}&0&{Q_{{2}}}^{4}
\\\noalign{\medskip}-1&{Q_{{3}}}^{2}&0\end {array} \right| , \left| 
\begin {array}{ccc} {Q_{{1}}}^{7}&{Q_{{1}}}^{3}&{Q_{{1}}}^{5}
\\\noalign{\medskip}{Q_{{2}}}^{6}&{Q_{{2}}}^{2}&0\\\noalign{\medskip}{
Q_{{3}}}^{5}&0&{Q_{{3}}}^{3}\end {array} \right|
\end{equation}
Sometimes it is helpful to use the generating series for multilinear Hirota operators:
\begin{equation}
H_k \ f \cdot \dots \cdot f=H\left[\prod\limits_{j=1}^{k}f_j\left(x+\sum\limits_{m=1}^{k-1}y_m e^{jm\frac{2\pi i}{k}} \right) \right]\vline_{y_1=\dots=y_{k-1}=0}
\end{equation}
Quadrilinear operator can be derived as follows:
\begin{equation}
Q^{l,m,n}= \frac{\partial^{i+j+k}}{\partial y_1^l\partial y_2^m \partial y_3^n}
(f \left( x+y_{{1}}+y_{{2}}+y_{{3}} \right) f \left( x+iy_{{1}}-y_{{2}}
-iy_{{3}} \right) \cdot
\end{equation}
\[
\cdot f \left( x-y_{{1}}+y_{{2}}-y_{{3}} \right) f \left( 
x-iy_{{1}}-y_{{2}}+iy_{{3}} \right))\vline_{y_1=y_2=y_3=0}
\]
Why roots of unity? Consider Taylor expansion of the product
\begin{equation}
f(x+c_1y)f(x+c_2y)f(x+c_3y)\vert_{y_1=y_2=y_3=0}
\end{equation}
Expanding the coefficients, we get
\begin{equation}
c_1= \left( c_{{3}}+c_{{2}}+c_{{1}} \right) {f}^{2}f_{{x}},
\end{equation}
\begin{equation}
c_2= \left(  \left( c_{{2}}+c_{{1}} \right) c_{{3}}+c_{{1}}c_{{2}}
 \right) f{f_{{x}}}^{2}+1/2\, \left( {c_{{3}}}^{2}+{c_{{2}}}^{2}+{c_{{
1}}}^{2} \right) {f}^{2}f_{{x,x}}.
\end{equation}
The coefficients are clearly homogeneous of weight $kp+s$ and satisfy the following property:
\begin{equation}
c_s(e^x \cdot f)=c_s(f)+\sum\limits_{k=0}^{s-1}a_kc_k(f)
\end{equation}
Equalling $c_1$ to zero implies all other coeffs to be gauge-invariant. When both of $c_1,c_2$ are zero,
only powers by modulo $3$ left in expansion, and we immediately get $\{T^i\}$ -- sequence of trilinear operators.

Identically zero condition:
\begin{equation}
\sum_{i=1}^{k-1}c_i\cdot\mathrm{binomial}(k-2,i)m_{i}\ \mathrm{mod}\ k \neq 0
\end{equation}
implies $H_{n,1}^{i_1}H_{n,2}^{i_2}\dots H_{n,n-1}^{i_{n-1}}\equiv 0$
($c_i=1$ if $k$ is prime). Example: $i-j\ \mathrm{mod}\ 3=0, i+2j-k\ \mathrm{mod}\ 4=0$.

Invariance condition:
\[
\left( H_k \left( {e}^{{x}^{i-1}}f \right)  \right) ^{m}={H_k}^{
m}f,0<i\leq m, \quad i=m \Leftrightarrow m | k
\]

Now we can obtain the $\varepsilon$-expansion of $k$-linear Hitora operator,
which is useful for finding solitonic or polynomial solutions. We use the differentiantion
rule
\[
H_k( f \cdot f \cdot ... \cdot 1) =\frac{d}{df}H_k( f \cdot f \cdot \dots \cdot f)
\]
For simplicity, we consider $k=3$.
\begin{equation}\label{TT*}
\mathcal{T}=\sum\limits_{i,j} T^iT^{*j},
\end{equation}
is any finite linear combination of $T$'s.
Substituting $f=\sum_i{}\varepsilon^if_i$ in \ref{TT*} ($\varepsilon$ is a formal parameter), we get the following
perturbation expansion:
\\ \ \\
\begin{tabular}{l l}
$\varepsilon$ & $Lf_1$ \\
$\varepsilon^2$ & $(\mathcal{T}f_1\cdot{}f_1\cdot{}f_1)'_{f_1}+Lf_2$ \\
$\varepsilon^3$ & $(\mathcal{T}f_1\cdot{}f_1\cdot{}f_1)+\sum\limits_{i=1}^{n}(\mathcal{T}f_1\cdot{}f_1\cdot{}f_1)''_{f_1^{(0)}f_1^{(1)}}f_2^{(i)}+Lf_3$ \\
$\varepsilon^4$ & $\sum\limits_{i=1}^{n}(\mathcal{T}f_1\cdot{}f_1\cdot{}f_1)'_{f_1^{(1)}}f_2^{(i)}+
\sum\limits_{i=1}^{n}(\mathcal{T}f_1\cdot{}f_1\cdot{}f_1)''_{f_1^{(0)}f_1^{(1)}}f_3^{(i)}+Lf_4$ \\
$\varepsilon^5$ & $\sum\limits_{i=1}^{n}\sum\limits_{j=1}^{n}(\mathcal{T}f_1\cdot{}f_1\cdot{}f_1)''_{f_1^{(i)}f_1^{(j)}}f_3^{(i)}f_3^{(j)}+\sum\limits_{i=1}^{n}(\mathcal{T}f_1\cdot{}f_1\cdot{}f_1)'_{f_1^{(1)}}f_3^{(i)}+$ \\ & $+\sum\limits_{i=1}^{n}(\mathcal{T}f_1\cdot{}f_1\cdot{}f_1)''_{f_1^{(0)}f_1^{(1)}}f_4^{(i)}+Lf_5$ \\
$\vdots$
\end{tabular}

The following table shows multilinear representation of some known equations:

\begin{tabular}{|c|c|c|c|}

\hline
Equation & Bilinear & Trilinear & Quadrilinear\\

\hline
KdV 3 & $D_x(D_x^3 + D_t)$  & $2 T_x^2 T_x^{*2} + 3T_x^{*} T_t $ & $2 G1_x^2 G3_x^2 +  G1_x^2 G2_t^1$ \\
    & - & $ T_x^4 T_x^{*}+ T_x^{*} T_t  $ &  \\
     
KdV 5 & - & $20 T_x^3T_x^{*3} +7T_x^6 + 27 T_x^* T_t$ & \\
     & & - &  $ G2_x^1 G3_x^6+ 5G1_x^1 G2_x^3 G3_x^3  + 12 G1_x^2 G2_t^1 $ \\
     
\hline     
SK 5 & $D_x(D_x^5 + D_t)$ & $10T_x^3T_x^{*3}-T_x^6+9T_x^{*} T_t$ & \\
     & & $T_x^5 T_x^{*2} + T_x^2 T_t$ & \\

SK 7 & & $ 14 T_x^4 T_x^{*4} + 14 T_x^7 T_x^{*} + 27 T_x T_t$ & $3 G3_x^8 +
7 G1_x^2 G3_x^6 - 42 G1_x^2 G2_x^4 G3_x^2 + 
64 G1_x^2 G2_t^1$ \\
     & & - & $7 G1_x^2 G2_x^3 G3_x^4  +  G1_x^1 G2_x^1 G3_x^7 + 16 G1_x^2 G2_t^1$ \\

\hline
``new'' 7 & $D_x(D_x^7+D_t)$ & $35 T_x^4 T_x^{*4} - 4 T_x^{7} T_x^{*}  + 27  T_x T_t^{*}$ &  \\
     &  & - & $7 G2_x^4 G3_x^{*4} -  G1_x G2_x^{2} G3_x^{5}  +12 G1_x^2 G2_t^1$ \\

\hline
``tail'' 9 & $D_x(D_x^9+D_t)$ & ${14} T_x^5 T_x^{*5} - {5}T_x^8 T_x^{*2} + 9 T_x T_t^{*}$ & \\

& & & \\

\hline
\end{tabular}

In the next section we investigate two trilinear integrable equations, considered in \cite{Hiet1}, and 
we show that they are representable as systems of bilinear equations. For some information on Painleve test,
Fuchs indicies and other integrability features we refer reader to \cite{Gor1}.

\section{J8 equation}
\begin{equation}\label{J8}
\left(4T_x^7T_x^*+5T_x^4T_x^{*4}+\alpha(20T_x^3T_x^{*3}+7T_x^6)+9\beta{}T_x^2T_x^*T_t^*+\frac{9\alpha\beta}{2}T_xT_t^*\right)f\cdot f\cdot f=0
\end{equation}
This equation passes Painleve test and has 3SS for arbitary $\alpha$,$\beta$.
Fuchs index
\begin{equation}
 \left( \begin {array}{ccccccc} -1&2&3&4&8&9&10\\\noalign{\medskip}-2&
-1&3&5&8&10&12\\\noalign{\medskip}-7&-5&-1&8&10&12&18\end {array}
 \right) 
\end{equation}
For $f(x)=1+e^{p_ix-q_it}$ dispersion law is
\begin{equation}
q_{{i}}=2\,{\frac {{p_{{i}}}^{5} \left( 3\,\alpha+{p_{{i}}}^{2}
 \right) }{\beta\, \left( 2\,{p_{{i}}}^{2}+\alpha \right) }}
\end{equation}
and for 2SS $f(x)=1+e^{p_1x-q_1t}+e^{p_2x-q_2t}+A_{1,2}e^{p_1x-q_1t}e^{p_2x-q_2t}$ we have
\begin{equation}
A_{1,2}={\frac { \left( p_{{1}}-p_{{2}} \right) ^{2} \left( 2\,{p_{{1}}}^{2}-2
\,p_{{2}}p_{{1}}+3\,\alpha+2\,{p_{{2}}}^{2} \right) }{ \left( p_{{1}}+
p_{{2}} \right) ^{2} \left( 2\,{p_{{1}}}^{2}+2\,p_{{2}}p_{{1}}+3\,
\alpha+2\,{p_{{2}}}^{2} \right) }}
\end{equation}
so $\alpha=0$ gives SK5 solution. Next we take $\alpha=0,\beta=1$ and show that (\ref{J8}) can be derived from SK5.

SK5 equation:
\begin{equation}
\left(u_{{x,x,x,x}}+30\,u_{{x,x}}u+60\,{u}^{3}_{{x}}\right)_x+u_{{t}}=0
\end{equation}
also can be written in a bilinear form with $u=\partial_{x,x}\log(f)$:
\begin{equation}\label{SK5b}
D_x(D_x^5+D_t)f\cdot f=0
\end{equation}
which is equivalent to
\begin{equation}
\left(-T_x^6+10T_x^3T_x^{*3}+9T_xT_t^*\right)f\cdot f \cdot f=0
\end{equation}
Now we compare two forms:
\begin{equation}\label{J81}
J8=T_xT_x{^*}\left(4T_x^6+5T_x^3T_x^{*3}+9T_xT_t^*\right)f\cdot f\cdot f
\end{equation}
and bilinear form of SK5 (\ref{SK5b}). J8 evaluates to
\\ \ \\
\begin{math}
J8=( f_{{t,x,x,x}}+f_{{x,x,x,x,x,x,x,x}} ) {f}^{2}+ ( -f
_{{x,x,x}}f_{{t}}-5\,{f_{{x,x,x,x}}}^{2}+3\,f_{{x,x}}f_{{t,x}}+4\,f_{{
x,x,x,x,x}}f_{{x,x,x}}+8\,f_{{x,x,x,x,x,x}}f_{{x,x}}-8\,f_{{x,x,x,x,x,
x,x}}f_{{x}}-3\,f_{{t,x,x}}f_{{x}} ) f-40\,f_{{x,x}}{f_{{x,x,x}}
}^{2}+20\,f_{{x,x,x,x,x,x}}{f_{{x}}}^{2}+60\,f_{{x,x,x,x}}{f_{{x,x}}}^
{2}+20\,f_{{x}}f_{{x,x,x,x}}f_{{x,x,x}}-60\,f_{{x}}f_{{x,x,x,x,x}}f_{{
x,x}}
\end{math}
\\ \ \\
Mult. (\ref{SK5b}) by $f$ and diff. twice, we get
\\ \ \\
\begin{math}
(SK5\cdot f)_{x,x}=
 ( f_{{t,x,x,x}}+f_{{x,x,x,x,x,x,x,x}} ) {f}^{2}+ ( -2
\,f_{{x,x,x,x,x,x,x}}f_{{x}}+5\,f_{{x,x,x,x,x,x}}f_{{x,x}}+4\,f_{{x,x,
x,x,x}}f_{{x,x,x}}+3\,f_{{t,x,x}}f_{{x}}-f_{{x,x,x}}f_{{t}}-5\,{f_{{x,
x,x,x}}}^{2} ) f-10\,f_{{x,x}}{f_{{x,x,x}}}^{2}+12\,f_{{x}}f_{{x
,x,x,x,x}}f_{{x,x}}-10\,f_{{x}}f_{{x,x,x,x}}f_{{x,x,x}}-3\,f_{{x}}f_{{
x,x}}f_{{t}}-10\,f_{{x,x,x,x,x,x}}{f_{{x}}}^{2}+15\,f_{{x,x,x,x}}{f_{{
x,x}}}^{2}
\end{math}
\\ \ \\
Direct computation shows that
\begin{equation}
J8-(SK5\cdot f)_{x,x}=3(SK5\cdot f_x)_{x}+R(x)
\end{equation}
where
\begin{equation}
R(x)=-9\, \left( f_{{x,x,x,x,x,x,x}}f-5\,f_{{x,x,x,x,x,x}}f_{{x}}+9\,f_{{x,x
,x,x,x}}f_{{x,x}}-5\,f_{{x,x,x,x}}f_{{x,x,x}}+f_{{t,x,x}}f-f_{{x,x}}f_
{{t}} \right) f_{{x}}
\end{equation}
which is nothing but $(SK5)_x$ mult. by $-9f_x$. So
\begin{equation}
J8=(SK5\cdot f)_{x,x}+3(SK5\cdot f_x)_{x}-9f_x(SK5)_x=f(SK5)_{x,x}-4D_x (SK5\cdot f)
\end{equation}

\section{J10 equation} 
\begin{equation}\label{J10}
\left(5T_x^8T_x^{*2}+4T_x^5T_x^{*5}+\alpha(4T_x^7T_x^{*}+5T_x^4T_x^{*4})+\beta(20T_x^3T_x^{*2}T_t^{*}+7T_x^5T_t)+
6\alpha\beta{}T_x^2 T_x^*T_t^*+\frac{3\beta^2}{2}T_tT_t^*\right)f\cdot f\cdot f=0
\end{equation}
Fuchs indices:
\begin{equation}
 \left( \begin {array}{cccccccc} -1&2&3&4&5&6&12&13
\\\noalign{\medskip}-2&-1&2&3&5&9&12&16\\\noalign{\medskip}-3&-2&-1&4&
5&11&12&18\\\noalign{\medskip}-13&-11&-7&-1&12&16&18&30\end {array}
 \right)
 \end{equation}
Dispersion law:
\begin{equation}
q_{{i}}={\frac { \left( 2\,\alpha+9\,{p_{{i}}}^{2}+\pm \left( \sqrt {4\,{\alpha}^{2}+30\,\alpha\,{p_{{i}}}^{2}+75\,{p_{{i}}
}^{4}} \right)  \right) {p_{{i}}}^{3}}{\beta}}
\end{equation}
Again, 2SS exists for arbitary $\alpha$, $\beta$, but its coeff. $A_{1,2}$ is very large. 3SS also exists.
In order to exclude J8 terms from (\ref{J10}), we assume $\alpha=0$:
\begin{equation}\label{J10}
J10=\left(5T_x^8T_x^{*2}+4T_x^5T_x^{*5}+\beta(20T_x^3T_x^{*2}T_t^{*}+7T_x^5T_t)+\frac{3\beta^2}{2}T_tT_t^*\right)f\cdot f\cdot f=0
\end{equation}
2SS for J10 is given by:
\begin{equation}
A_{1,2}=-{\frac { \left( {p_{{2}}}^{2}+\sqrt {3}p_{{1}}p_{{2}}+{p_{{1}}}^{2}
 \right)  \left( {p_{{2}}}^{2}-p_{{1}}p_{{2}}+{p_{{1}}}^{2} \right) 
 \left( -p_{{2}}+p_{{1}} \right) ^{2}}{ \left( -{p_{{2}}}^{2}+\sqrt {3
}p_{{1}}p_{{2}}-{p_{{1}}}^{2} \right)  \left( {p_{{2}}}^{2}+p_{{1}}p_{
{2}}+{p_{{1}}}^{2} \right)  \left( p_{{2}}+p_{{1}} \right) ^{2}}}
\end{equation}
This will lead us to associated bilinear form given by:
\begin{equation}
 \left( q-{\frac { \left( 9-5\,\sqrt {3} \right) {p}^{5}}{\beta}}
 \right)  \left( q-{\frac { \left( 9+5\,\sqrt {3} \right) {p}^{5}}{
\beta}} \right) 
\end{equation}
Corr. bilinear operator is defined uniquely:
\begin{equation}\label{bil}
B1=\left(\frac{6}{\beta^2} D_x^{10}+\frac{18}{\beta} D_x^5 D_t +D_t^2\right)f\cdot f
\end{equation}
This eq. (\ref{bil}) has 2SS, but not 3SS. Its coeff.:
\begin{equation}
A_{1,2}={\frac { \left( {p_{{2}}}^{2}-p_{{1}}p_{{2}}+{p_{{1}}}^{2} \right) 
 \left( 2\,{p_{{1}}}^{2}-3\,p_{{1}}p_{{2}}+\sqrt {3}p_{{1}}p_{{2}}+2\,
{p_{{2}}}^{2} \right)  \left( {p_{{1}}}^{2}+\sqrt {3}p_{{1}}p_{{2}}+{p
_{{2}}}^{2} \right)  \left( p_{{1}}-p_{{2}} \right) ^{2}}{ \left( -{p_
{{1}}}^{2}+\sqrt {3}p_{{1}}p_{{2}}-{p_{{2}}}^{2} \right)  \left( -2\,{
p_{{1}}}^{2}-3\,p_{{1}}p_{{2}}+\sqrt {3}p_{{1}}p_{{2}}-2\,{p_{{2}}}^{2
} \right)  \left( {p_{{2}}}^{2}+p_{{1}}p_{{2}}+{p_{{1}}}^{2} \right) 
 \left( p_{{1}}+p_{{2}} \right) ^{2}}}
\end{equation}
Now we are going to represent (\ref{J10}) as a system of bilinear equations with (\ref{bil})
plus some additional relations. At first we rewrite (\ref{bil}) in trilinear form.
\begin{eqnarray}
(f\cdot D_t^2)= \frac{2}{3}T_tT_t^*\\
(f\cdot D_x^6)=\frac{2}{27}(-T_x^6+10T_x^3T_x^{*3})\\
(f\cdot D_x^5D_t)=\frac{2}{27}(-T_x^5T_t+10T_x^3T_x^{*2}T_t^*)\\
(f\cdot D_x^{10})=\frac{2}{27}(-5T_x^8T_x^{*2}+14T_x^5T_x^{*5})
\end{eqnarray}
Dividing by lead coeff., we get
\begin{equation}
f\cdot B1=-5T_x^8T_x^{*2}+14T_x^5T_x^{*5}+3\beta(-T_x^5T_t+10T_x^3T_x^{*2}T_t^*)+\frac{3\beta^2}{2}T_tT_t^*
\end{equation}
So thier difference is symmetrical:
\begin{equation}
J10-(f\cdot B1)=10(\beta(T_x^5T_t-T_x^3T_x^{*2}T_t^*)+(T_x^8T_x^{*2}-T_x^5T_x^{*5}))
\end{equation}
Note here coeffs summ to zero!
Here both terms $T_x^5T_t-T_x^3T_x^{*2}T_t^*,T_x^8T_x^{*2}-T_x^5T_x^{*5}$ are zero for $f(x,t)=1+e^{px-qx}$
for any $p,q$. Subs. $f(x,t)=1+e^{p_1x-q_1x}+e^{p_2x-q_2x}$ in the second term and factorizing, we get
\begin{equation}
-3\,{p_{{1}}}^{2}{{\rm e}^{p_{{1}}x-q_{{1}}t+p_{{2}}x-q_{{2}}t}}{p_{{2
}}}^{2} \left( -p_{{2}}+p_{{1}} \right) ^{2} \left( {p_{{2}}}^{2}-p_{{
1}}p_{{2}}+{p_{{1}}}^{2} \right) ^{2}
\end{equation}
This is 2SS numer for the new7 eq. One can check that
\begin{equation}
(T_x^8T_x^{*2}-T_x^5T_x^{*5})f\cdot f\cdot f\equiv c_1D_x^3(D_{\tau}+D_x^7)f \cdot f,
\end{equation}
with
\begin{equation}
D_x(D_{\tau}+D_x^7)f\cdot f=0
\end{equation}
(it is easier with $u(x,t,tau)$ variable)

Another term is similar to Lax5 and it splits to
\begin{equation}
(T_x^5T_t-T_x^3T_x^{*2}T_t^*)f\cdot f\cdot f\equiv c_2D_x^3(D_{\tau'}+D_x^2D_t)f \cdot f,
\end{equation}
with
\begin{equation}
D_x(D_{\tau'}+D_x^2D_t)f \cdot f=0
\end{equation}
So finally J10 is given by
\begin{eqnarray}
\left(\frac{6}{\beta^2} D_x^{10}+\frac{18}{\beta} D_x^5 D_t+D_t^2+c_1D_x^3(D_{\tau}+D_x^7)+c_2D_x^3(D_{\tau'}+D_x^2D_t)\right)f\cdot f=0 \\
D_x(D_{\tau}+D_x^7)f\cdot f=0 \\
D_x(D_{\tau'}+D_x^2D_t)f\cdot f=0
\end{eqnarray}

Summarizing all the above, such equations can be considered as the $l$-linear dominant part of order $l+m-1$ parametrized by bilinear relations.
\begin{eqnarray}\label{bil_pairs}
D_x^l(D_{\tau}+D_x^m)(f\cdot f)+\dots=theequation, \\
D_x(D_{\tau}+D_x^m)(f\cdot f)=0
\end{eqnarray}
We will denote it as $(D^l,D^m)$.
The following table represents $l,m$ bilinear pairs of lowest orders:

\begin{tabular}{|c|c|c|c|c|c|c|c|c|c|}
\hline
$m,l$ & 3 & 5 & 7 & 9 & 11 \\
\hline
3 & Lax5 & Lax7 & - & - & - \\
\hline
5 & SK7/H8 & H10 & - & - & - \\
\hline
7 & J10 & - & new13 & - & - \\
\hline
9 & - & - & - & - & - \\
\hline
11 & - & - & - & - & - \\
\hline
\end{tabular}

\section{Higher order equations}

In this section we represent some results on higher order generalizations of J8,J10 equations.
Firstly, we formulate the principle.

\textbf{Property. 1.}
\emph{For $D^n$ Hirota operator fuchs array is symmetric over negative indices}

\textbf{Property. 2.}
\emph{If equations $E_1,E_2,\dots$ are in a KP-type hierarchy with $deg(E_1)<deg(E_2)<\dots$, then}
\begin{equation}
s_1\subset s_2 \subset \dots,
\end{equation}
\emph{where $s_i$ is a triangular sub-array of negative fuchs indices of $E_i$}

Using these properties, we can build  higher order analogues of J10 equation. 
The equations should be organized in the way (in a manner similar to the KP, BKP hierarchies \cite{JM1}):
\[
D^{10}+(D^3,D^7)=J10,
\]
\[
D^{12}+(D^5,D^7)+(D^3,J10)=J12,
\]
\[
D^{14}+(D^7,D^7)+(D^5,J10)+(D^3,J12)=J14,
\]
\[
\dots
\]
Namely, consider degree-homogeneous linear combination (see \ref{bil_pairs})
\begin{equation}
D^{12}+k_1(D^5,D^7)+k_2(D^3,J10)
\end{equation}
We can obtain the coefficients $k_i$ resolving all fuchs indices one by one, so that
\begin{equation}
F= \left( \begin {array}{cccccc} - 1.0& 1.0& 2.0& 3.0& 4.0& \dots\\
\noalign{\medskip}- 2.0&- 1.0& 1.0& 2.0&
 3.0\\
 \noalign{\medskip}- 3.0&- 2.0&- 1.0& 1.0& 4.0& \dots \\
\noalign{\medskip}- 13.0&- 11.0&- 7.0&- 1.0& 0.36985419&\dots \\
\noalign{\medskip}-
 13.76327806&- 11.56856256&- 7.247114834& - 0.3858031569& -1.0& \dots
 \end {array}
 \right)
\end{equation}
The corresponding values are
\begin{equation}
k_1=27,k_2=-28.
\end{equation}
However, due to the absence of irrational indices, this equation might not be integrable.
But it is closely related to the determinat-like 12th order extension of J10 (\ref{J12}), which is already integrable!
\begin{equation}\label{J12}
J12=-18\,P_{{[1,0,5,6]}}+4\,P_{{[1,5,0,6]}}+5\,P_{{[1,10,0,1]}}-80\,P_{{[3
,3,3,3]}}-35\,P_{{[3,4,1,4]}}+
\end{equation}
\[
+10\,P_{{[4,0,1,7]}}+60\,P_{{[4,2,2,4]}}+
30\,P_{{[4,3,0,5]}}+24\,P_{{[5,1,1,5]}}
\]
Expanding all pentalinear operators in \ref{J12}, we get:
\[
J12=\left( 2\,f_{{8\,x}}f_{{4\,x}}-2\,f_{{7\,x}}f_{{5\,x}}-2\,f_{{9\,x}}f
_{{3\,x}}+{f_{{6\,x}}}^{2}+f_{{10\,x}}f_{{2\,x}} \right) {f}^{3}+\left( 28\,f_{{6\,x}}{f_{{3\,x}}}^{2}-f_{{10\,x}}{f_{{x}}}^{2}+9\,f_{
{8\,x}}{f_{{2\,x}}}^{2}+\right.
\]
\[
+
\left.2\,f_{{5\,x}}f_{{6\,x}}f_{{x}}+6\,{f_{{5\,x}}}
^{2}f_{{2\,x}}+10\,f_{{8\,x}}f_{{x}}f_{{3\,x}}-4\,f_{{9\,x}}f_{{x}}f_{
{2\,x}}-30\,f_{{5\,x}}f_{{3\,x}}f_{{4\,x}}-6\,f_{{7\,x}}f_{{4\,x}}f_{{
x}}+\right.
\]
\[
+\left.10\,{f_{{4\,x}}}^{3}+4\,f_{{6\,x}}f_{{2\,x}}f_{{4\,x}}-28\,f_{{7\,
x}}f_{{2\,x}}f_{{3\,x}} \right) {f}^{2}+\dots
\]

\section{Summary}
As we have shown, multilinear Hirota equations arise from the KP-type bilinear hierarchies,
i.e. they can be expanded as $n$-tuples of bilinear equations, each of them is contained in a hierarchy
(we recall here that such hierarchies are intimately linked with the affine Kac-Moody algebras).
This fact follows from the Lie-algebraic structure of these equations.
Thus, gauge-invariant multilinear operators of higher order can be useful
for describing such higher reductions. However, when the reduction is done, total dimension is not greater
than (2+1). It is also an interesting fact that J10 integrable equation consists of 3 bilinear equations, one of which is not integrable in the ``default'' time-variables. But its (2+1) extension is truly integrable
(can be obtained by adding BKP even terms to J10 and assuming a solution in form of diagonal Schur function).
We also find J12 analogue as the determinant-like equation (it is non-evolutionary and dispersionless).
And it remains unclear if there are connections with hyperbolic Kac-Moody algebras. The reasonable approach is as follows.
One should extend the space of symmetric functions (using some natural deformations) and define an operator similar to Hirota derivative, which eigenvectors are connected with $U_q(\mathfrak{sl}_2(\mathbb{C}))$. This is of further development.



\begin{thebibliography}{1}
\bibitem{Ath1}
C. Athorne Algebraic invariants and generalized Hirota derivatives, Physics Letters A, vol. 256 n. 1 p. 20, 1999
\bibitem{Ath2}
M. England, C. Athorne Building Abelian Functions with Generalised Baker–Hirota Operators, SIGMA 8, 2012, 037
\bibitem{Gor1}
A. Goriely Integrability and Nonintegrability of Dynamical Systems, Advenced Series in Nonlinear Dynamics vol. 19, WSP, 2001
\bibitem{Hiet1}
J. Hietarinta, B. Grammaticos Integrable trilinear PDE's, arXiv preprint solv-int/9411003, 1994
\bibitem{Hiet2}
J. Hietarinta Hirota’s bilinear method and soliton solutions, Physics AUC, vol. 15 (part I), 31-37, 2005
\bibitem{JM1}
M. Jimbo, T. Miwa Solitons and Infinite Dimensional Lie Algebras, Publ RIMS, Kyoto Univ. 19, 1983, 943-1001
\bibitem{Lu1}
S. Lu On Soliton Equations of Exceptional Type, Journal of Algebra 166, 611-629, 1994
\end{thebibliography}
\end{document}